\documentstyle[11pt,newpasp,twoside,epsf]{article}
\begin{document}

%%%%%%%%%%%%%%%%%% title %%%%%%%%%%%%%%%%%%%%%%%%%%%%%%%%%%%%%%%%

\title {\bf Dust Continuum Observations of Massive Star Forming Regions}
\author {Kaisa E. Mueller, Yancy L. Shirley, Neal J. Evans II, and Heather R. Jacobson}
\affil{Department of Astronomy, The University of Texas at Austin, RLM 15.308, C1400,
       Austin, Texas 78712--1083}

%%%%%%%%%%%%%%%%%% abstract %%%%%%%%%%%%%%%%%%%%%%%%%%%%%%%%%%%%%%%%
 
\begin{abstract}
We have observed 51 high mass star forming cores associated
with water masers at 350\micron.  The spectral energy distributions (SEDs) and dust continuum normalized radial intensity profiles were modeled for 28 sources using a one-dimensional dust radiative transfer code assuming a power law density distribution in the envelope $n = n_0 (r/r_0)^{-p}$. The best fit density power law exponent, $p$, ranged from 1.0 to 2.5 with a mean value of $p = 1.72$. We determined the dust masses for the modeled cores and found a mean mass of 209 M$_{\sun}$.
\end{abstract}

%%%%%%%%%%%%%%%%%% 1. Main Text %%%%%%%%%%%%%%%%%%%%%%%%%%%%%%%%%%%%%%%%

\section{Observations}
More than fifty high mass star forming cores were observed with SHARC (Submillimeter High
Angular Resolution Camera; Hunter, Benford, \& Serabyn 1996) during 5 nights in 1997 (December 21 and 22) and 1998 (July 15, 23, and 25) on the 10.4m Caltech Submillimeter Telescope. SHARC is a 1D bolometer array with a FWHM beam size of 11\arcsec.  The objects were selected from the sample of Plume, Jaffe, \& Evans (1992) of massive star forming cores associated with water masers. We have plotted the 350\micron\ contour maps and normalized radial intensity profiles for each source. Figure 1 shows the 350\micron\ map and the radial profile with the best-fit model for M8E.  

%%%%%%%%%%%%%%%%%%%Figure 1%%%%%%%%%%%%%%%%%%%%%%%
\begin{figure}
\plotfiddle{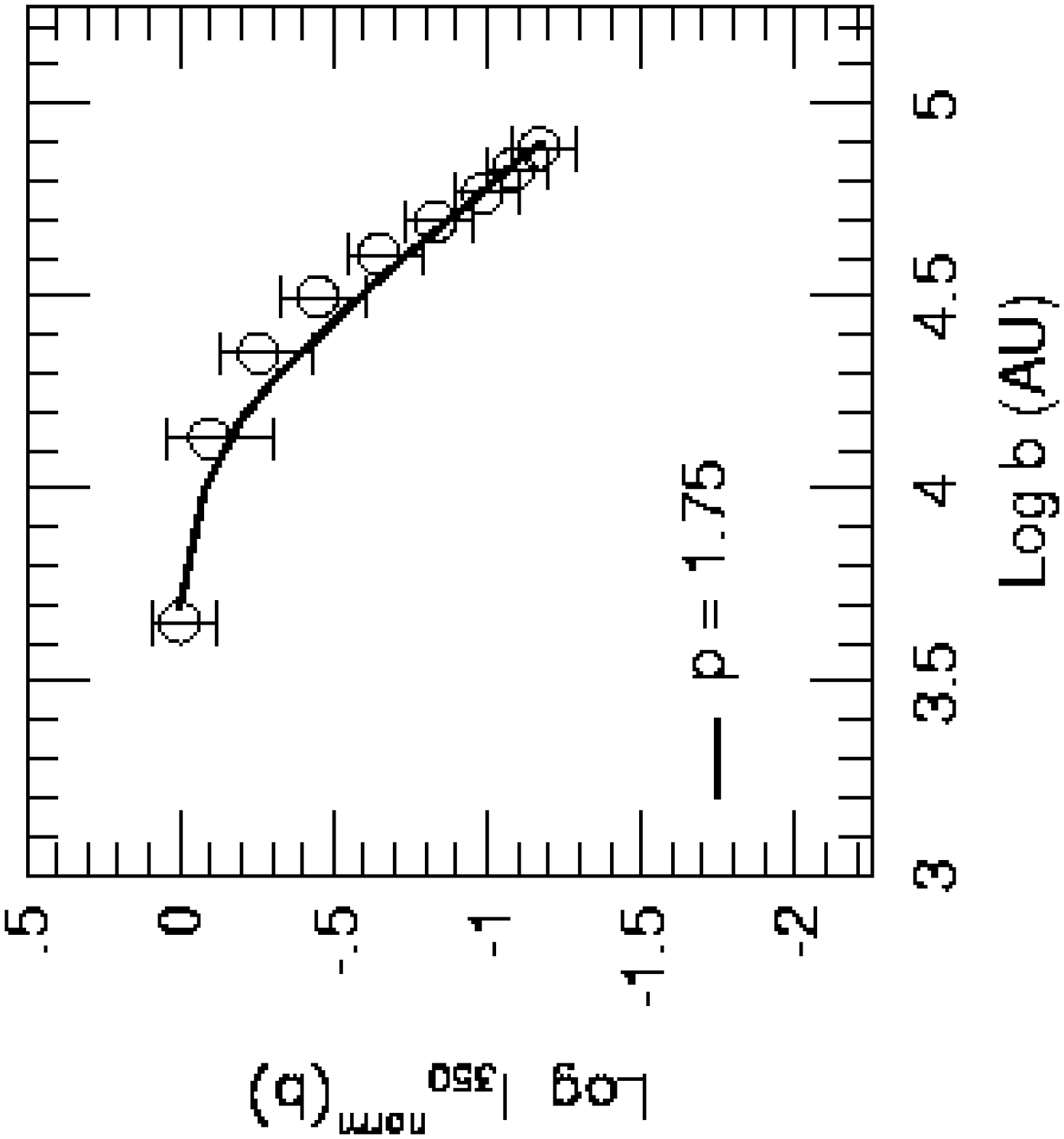}{145pt}{270}{32}{32}{-15}{155}
\plotfiddle{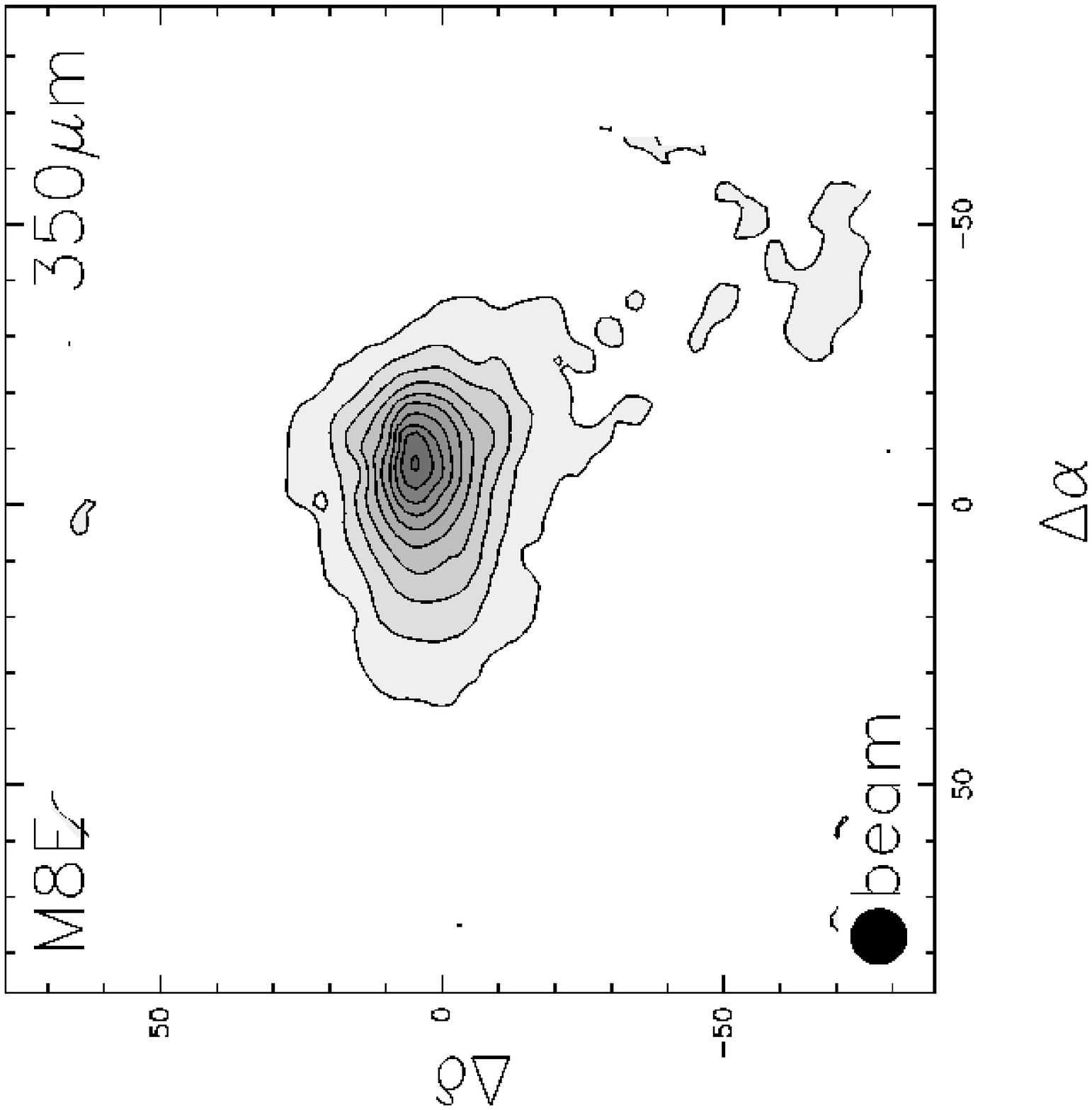}{0pt}{270}{30}{30}{-190}{175}
\caption{350\micron\ map (contours are 10\% of the peak or 5.4$\sigma$) and normalized radial profile of M8E ($18^h 01^m 49.1^s,  -24^{\circ} 26\arcmin\ 57\arcsec$). Solid line plotted with the radial profile is the best-fit model $p = 1.75$, $\chi^2$ = 0.52.}
\end{figure}
%%%%%%%%%%%%%%%%%%%%%%%%%%%%%%%%%%%%%%%%%%%%%%%%%

\section{Models}
The radial intensity profiles and SEDs were modeled using a one-dimensional dust continuum radiative transfer code by Egan, Leung, and Spagna (1988) and an observation simulation code (Evans et al. 2001).  We assumed a power law density distribution of the form $n \propto r^{-p}$, for $p = 0.5 - 2.5$. For each of our models, we adjusted the input density so that the model flux at 350\micron\ matched the observed flux and adjusted the internal luminosity to match the observed bolometric luminosity. We adopted dust opacities from Ossenkopf \& Henning (1994, column 5) that are calculated for aggregated dust grains with ice mantles as our standard. The fit of the models with the observations was quantified by calculation of a reduced chi-squared, $\chi^2$.

\section{Results} 

The internal luminosities of our sources ranged from 10$^3$ to 10$^6$ L$_{\sun}$. For all the modeled sources, we have calculated an integrated mass using the best-fit density power law.  The mean mass within the half-power radius is 209 M$_{\sun}$. (See Evans et al. and Shirley et al. in these proceedings for the mean virial mass as determined from CS observations.)  

Averaged over 28 sources with models, $\langle p\rangle = 1.72 \pm 0.37$.  We were unable to reliably model sources with small angular sizes or double peaks, and those cores are not included in our statistics.  The mean power law is significantly steeper (about 0.4) than the average density law exponent found by van der Tak et al. (2000) in a study of 14 massive star formation regions.  Our models were convolved with the observed beam profile rather than a gaussian as used by van der Tak et al. (2000), accounting for the difference.

%%%%%%%%%%%%%%%%%%%%%% REFERENCES %%%%%%%%%%%%%%%%%%%%%%%%%%%%%%%%%%%%


\begin{references}

\reference{} Egan, M. P., Leung, C. M., \& Spagna, G. R. 1988, Comput. Phys. Comm., 48, 857
\reference{} Evans II, N. J., Rawlings, J. M. C., Shirley, Y. L., \& Mundy, L. G. 2001, \apj, 557, 193
\reference{} Hunter, T. R., Benford, D. J., \& Serabyn, E. 1996, \pasp, 108, 1042
\reference{} Ossenkopf, V., \& Henning, T. 1994, \aap, 291, 943
\reference{} Plume R., Jaffe, D. T., \& Evans II, N. J. 1992, \apjs, 78, 505
\reference{} van der Tak, F. F. S., van Dishoeck E. F., Evans II, N. J., \& Blake, G.A., 2000 \apj, 537, 283


\end{references}
\end{document}